\documentclass[aps,12pt,letterpaper,amsmath,showpacs,preprint]{revtex4}
\usepackage{graphicx}
\usepackage{epstopdf}
\usepackage{bm}
\usepackage[pdftex]{color}

\begin{document}
\date{\today}
\title{Anomalous Rabi oscillations in multilevel quantum systems}

\author{B. Y. Chang}
\affiliation{School of Chemistry (BK21), Seoul National University, Seoul 151-747, Republic of Korea}
\author{I. R. Sola}
\affiliation{Departamento de Qu\'imica F\'isica I, Universidad Complutense, 28040 Madrid, Spain}
\email{isola@quim.ucm.es}
\author{Vladimir~S.~Malinovsky}
\affiliation{US Army Research Laboratory, Adelphi, MD 20783}
\email{vladimir.s.malinovsky.civ@mail.mil}
\begin{abstract}
We show that the excitation probability of a state within a manifold of levels undergoes Rabi oscillations with frequency determined by the energy difference between the states and not by the pulse area for sufficiently strong pulses. The observed dynamics can be used as a procedure for robust state preparation as an alternative to adiabatic passage and as a useful spectroscopic method.
\end{abstract}


\maketitle 

Coherent excitation is a fundamental step for quantum state preparation, underlying a variety of implementations in quantum information and quantum control of atoms,  molecules and nano-devices. In general, quantum systems have complex structures, however, under certain conditions, their dynamics can be well described by just a few separated energy levels~\cite{Eberly,Scully,Berman}. In this regard, the two-level structure (TLS) represents a particularly simple and well-adapted system to represent a qubit
or to describe elementary control processes~\cite{Chuang,Moshe}.

Under a constant field of amplitude $\epsilon_0$ and frequency $\omega$,  slightly
detuned from the resonance, so that $\Delta = E_2 - E_1 - \hbar\omega$,
the excitation probability of the TLS undergoes Rabi oscillations
\begin{equation}
P_2(t) = \frac{\Omega_0^2}{\Omega_\mathrm{eff}^2} \sin^2\left( 
\Omega_\mathrm{eff} t / 2 \right) \, ,
\end{equation}
where $\Omega_0 = \mu \epsilon_0$ is the Rabi frequency (in atomic units),
and the effective Rabi frequency, $\Omega_\mathrm{eff} = \sqrt{\Omega_0^2 + \Delta^2}$,
takes into account the effect of the detuning. The detuning
accelerates the rate of the population transfer and decreases the maximum transfer,
hampering the efficiency and robustness of the preparation process.
A general feature of detuning is to generate fast oscillating dynamical
phases that modulate (and reduce) the coherent transfer induced
by the Rabi frequency.

Logically, we may expect that increasing the system complexity will make more difficult to control the 
population dynamics, since more states will be involved and more effective detunings will  modulate the Rabi oscillations. Indeed, one of the main problems of building quantum machines ({\it e.g.} quantum computers) is the ability to first isolate TLS and then couple these structures in a controllable way. In this work, however, we show that there are very general classes of structures, more complex than the TLS, where the additional states help to enhance the robustness of the population transfer. It turns out that in the presence of a driving field, the system dressed states are characterized by isolating a ''two-level dressed substructure" (TLDS) where the dynamics occurs. Assuming constant fields for simplicity, the effective Rabi frequency driving the population dynamics within these two dressed states is of the form
\begin{equation}\label{ERFtlds}
\Omega_\mathrm{eff} = \sqrt{ a\delta^2 + b\Omega_0^2 -
\sqrt{a^2\delta^4 + b^2\Omega_0^2 + c \delta^2\Omega_0^2}} \, ,
\end{equation} 
where $\delta$ is a characteristic energy splitting of the system, and $a,b,c$ are some relevant parameters. Whereas the dynamics of the ordinary TLS is governed by the largest system frequency (the Rabi frequency or the detuning), in the TLDS behavior is just the opposite; the driving force is always the 
slowest system frequency, given by the Rabi frequency, $\Omega_0$, at low field intensities, and by the splitting frequency, $\delta$, at large intensities. This self-regulated behavior makes these systems 
promising candidates from which to build quantum machines, owing to the greater robustness of their coherent population dynamics. Note that, in this case, the speed of quantum operations (gates) is constrained by the energy splitting of the system and not by the strength of the external drive. 
We will refer to the oscillations induced by the effective Rabi frequency of the form of Eq.~(\ref{ERFtlds}) or its time-dependent generalizations as anomalous Rabi oscillations (ARO).
 
To demonstrate the ARO we consider a four-level tripod system comprised of three non-degenerate levels in the ground state coupled to a single excited state. A simple experimental realization could be the excitation from a $J=1$ atomic state to an excited $J=0$ state by the fields of a suitable polarization to allow all the couplings. Degeneracy of the ground sublevels  can be lifted by a strong magnetic field that creates Zeeman splittings, $\delta = g\mu_B B$, where $g$ is the Land\'e factor, $\mu_B$ the Bohr magneton, and $B$ the magnetic field. Here we are interested in the regime where the splittings are larger than the pulse bandwidth and we assume that any Zeeman sublevel of the ground state can be initially prepared. A similar behavior occurs when the $J=0$ is the initial state and the $J=1$ is the final state.

Expanding the wave function as $|\Psi(t)\rangle = \sum_{M=-1}^{1} a_M(t) |M\rangle + a_{0'}(t) |0'\rangle$,  where $M$ is the magnetic quantum number in the ground state and the prime indicates the excited state, and applying the rotating wave approximation (RWA), the Hamiltonian can be written as
\begin{equation} \label{Ham4}
{\sf H} = \sum_M \delta M |M\rangle \langle M| + \Delta |0'\rangle \langle 0'|
- \Omega(t)/2 \left[ \sum_M |M\rangle \langle 0'| + \mathrm{h.c.} \right] \, ,
\end{equation} 
where $\Delta = E_{0'} - \hbar\omega$ is the single-photon detuning, $\omega$ is the carrier frequency, $\Omega(t)=\Omega_0 \exp\{-(t-t_c)^2/(2\tau_0^2)\} $, $\Omega_0$ is the peak Rabi frequency, $\tau_0$ determines the pulse duration, and h.c. refers to the hermitian conjugate components. Choosing $\Delta = \delta M$ we can resonantly excite the $|0'\rangle$ state from any initial sublevel. Starting in the $|0\rangle$ sublevel, we obtain the most symmetrical arrangement when $\Delta = 0$. Figure~1 shows a sketch of the system including the couplings for the symmetrical arrangement (a) as well as one possible asymmetric configuration (b), where $\Delta = -\delta$. In the figure we also present the time-dependent energies of
the system dressed states for both configurations.

\begin{figure}
\includegraphics[width=12cm]{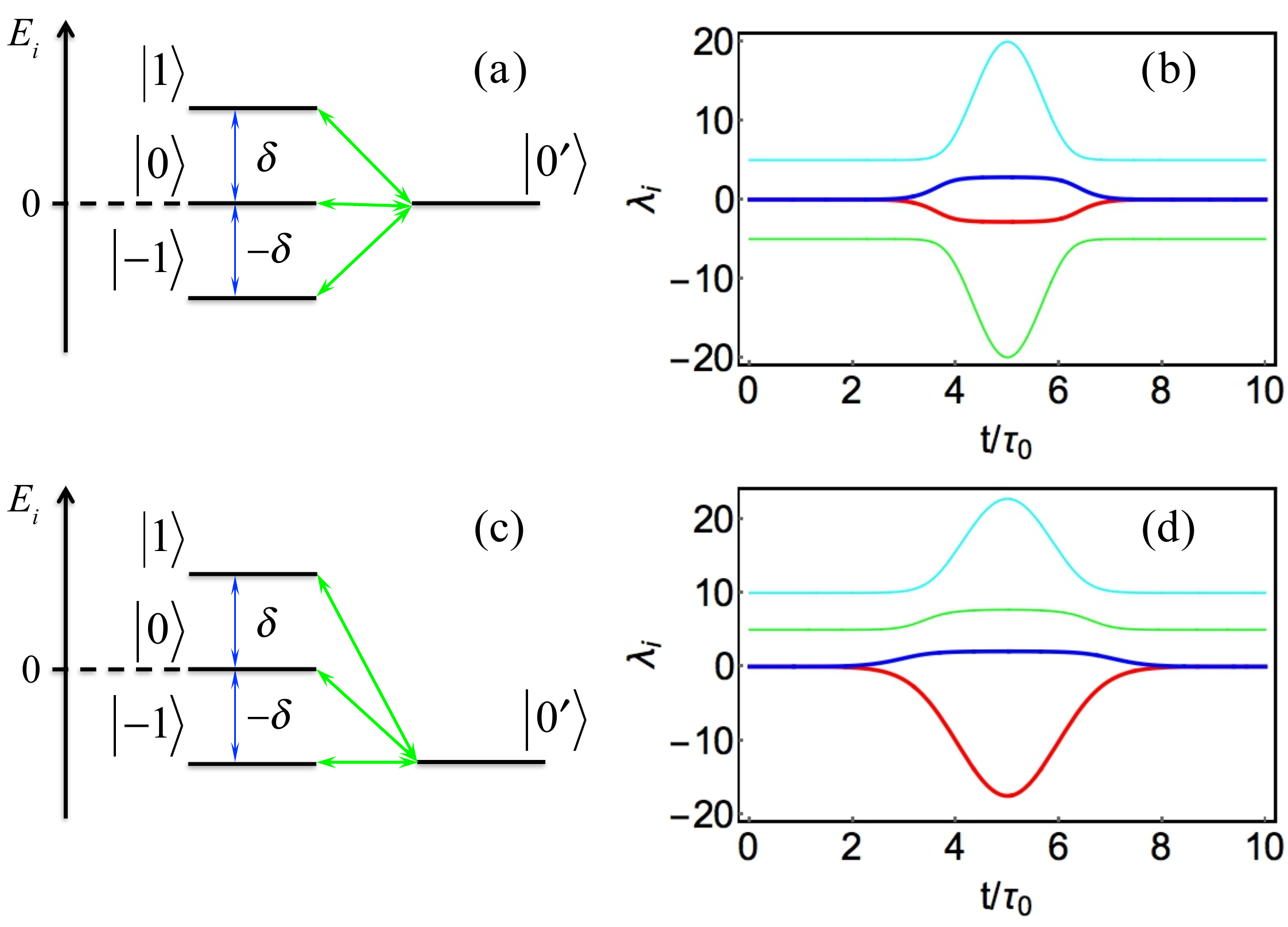}
\caption{(Color online) Tripod system in symmetric (a) and asymmetric (c) configurations and their
respective dressed states (b) and (d); $t_c=5\tau_0$, $\delta\tau_0=5$, $\Omega\tau_0=18$. The thicker  blue and red lines correspond to the populated dressed states.}
\label{Fig1} 
\end{figure}

First, we consider the symmetric configuration of the system. In the adiabatic limit, the system dynamics  can be described analytically. By diagonalizing the Hamiltonian, Eq.~(\ref{Ham4}), we obtain the time-dependent energies of the dressed states
\begin{align}
\lambda_{1,2}(t) &= \mp\frac{\sqrt{\delta^2+3 \chi(t)^2- D^2}}{\sqrt{2}} \, ,\\
\lambda_{3,4}(t) &= \mp\frac{\sqrt{\delta^2+3 \chi(t)^2+D^2}}{\sqrt{2}} \, ,
\end{align}
where $D^2=\sqrt{\delta^4+2 \chi^2(t) \delta^2+9 \chi^4(t)}$, $\chi(t) = \Omega(t)/2$. 
From these expressions we see two limiting cases: for the weak fields, when $\Omega_0 \ll \delta$, 
$\lambda_{1,2}(t) = \mp \chi(t)$, while for large pulse intensities, $\Omega_0 \gg \delta$, 
$\lambda_{1,2}(t) = \mp \delta/\sqrt{3}$. We also observe that the dressed states come in pairs: the distance between $\lambda_1(t)$ and $\lambda_2(t)$  is approximately bounded by the energy splitting due to avoided crossings with the other dressed states, and it remains small regardless of the pulse amplitude, while the distance between $\lambda_3(t)$ and $\lambda_4(t)$ increases following the Rabi frequency, $\Omega(t)$. 

For the initial condition $a_{0}(0)=1$, the wave function at the end of the pulse depends
only on the amplitudes $a_{0}$ and $a_{0'}$. Neglecting nonadiabatic couplings,
the analytic solution of the time-dependent Schr\"odinger equation (TDSE) is
\begin{align}
a_{0}(t) &= \frac{2 \chi^2(t) }{D \sqrt{D^2-\delta^2-\chi^2(t)}} \cos \left(\int_0^t \lambda_2(t^{\prime}) dt^{\prime}\right) \, ,\\
a_{0^{\prime}}(t) &=  \frac{2 i \delta \chi(t) }{D \sqrt{D^2-\delta^2+3\chi^2(t)}} \sin \left(\int_0^t \lambda_2(t^{\prime}) dt^{\prime}\right) \, .
\end{align}
At the end of the pulse ($t =T$) the target state population follows the area theorem $P = \sin^2({\cal A}/2)$ with ${\cal A} = 2\int_0^T dt \lambda_2(t)$, similar to the TLS dynamics. For small pulse intensities 
($\Omega_0 \ll \delta$), the area is proportional to the integral of the Rabi frequency,
\begin{equation}   
{\cal A} \approx \int_0^T dt \Omega(t) \left[ 1 - \frac{\Omega(t)^2}{4\delta^4}\right] \, .
\end{equation}
This is the expected result since the system is essentially a TLS.  There are small deviations
up to third order in the Rabi frequency (second term in the square braces) due to the presence of the $|\pm 1\rangle$ states. However, in the limit of large pulse intensities the area remains practically constant.
There is a coherent saturation effect and the yield of population transfer becomes much more robust as the field intensity increases, contrary to the standard TLS result~\cite{Eberly}. In practice this leads to very slow oscillations in the final state which are controlled by the Zeeman splitting, $\delta$.

\begin{figure}
\includegraphics[width=12cm]{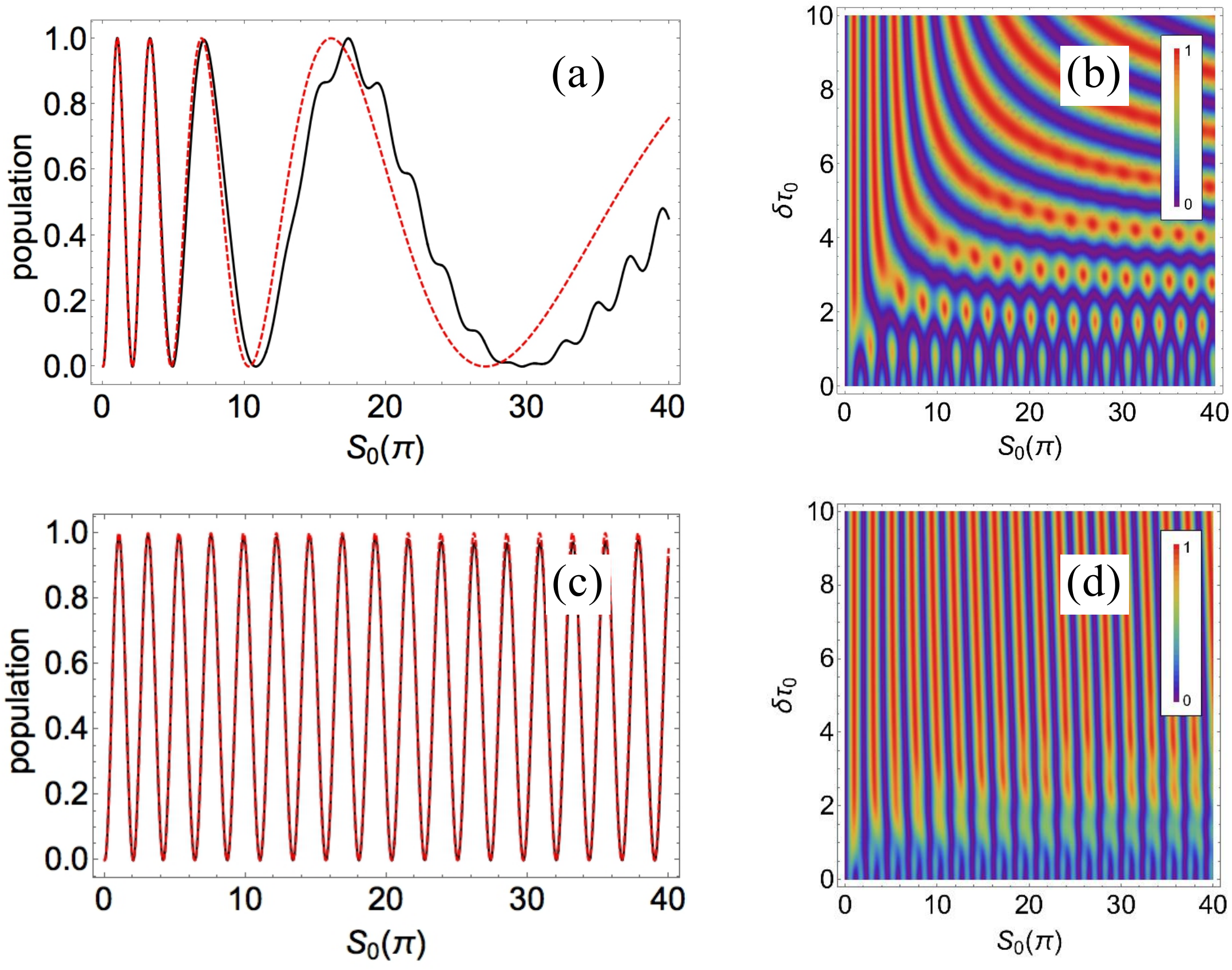}
\caption{(Color online) Coherent population transfer to the target state in the symmetric (a) and asymmetric (c) tripod schemes, as a function of the pulse area, for $\delta\tau_0 = 5$. The red dotted line represents the analytical results (neglecting nonadiabatic couplings) and the black solid line represents the numerical
solution of the TDSE. In (b) and (d) we show the target population at final time $T$ as a function of both pulse area $S_0$  and splitting  $\delta$.}
\label{Fig2}
\end{figure}

Figure~\ref{Fig2}(a) compares the final yield as a function of the pulse area in the adiabatic limit
(analytic results) with the result of numerical integration of the TDSE. We have used Gaussian pulses and we fixed $\delta\tau_0 = 5$. The agreement of the results shows that the non-adiabatic couplings are negligible until $\Omega_0$ is much larger than $\delta$, when the states $|\pm 1\rangle$ also become populated at final time. Surprisingly, in the TLDS the adiabaticity is damaged at large pulse intensities in contrast  to commonly accepted adiabatic criterion~\cite{Bergmann}. 

Figure~\ref{Fig2}(b) shows a density plot of the final yield of population transfer as a function of both 
pulse area, $S_0=\int_0^T \Omega(t) dt$,  and splitting,  $\delta$, calculated numerically. For not very large Rabi frequencies the final-time population oscillates between the initial and the target state as it is in the traditional TLS. The states $|\pm 1\rangle$ are only populated due to non-adiabatic couplings, leading first to some higher-frequency modulation or wiggles of the ARO and lower yields, and then to some enlargement of the oscillation period since part of the pulse energy is used to excite other Zeeman levels via Raman transitions. These effects are due to some population transfer near the crossings of $\lambda_{1,2}(t)$ with $\lambda_{3,4}(t)$ when $\Omega(t) \approx \delta$ at the beginning and end of the pulse. Hence the nonadiabatic effects are less important the larger the energy splitting is.

The presence of the ARO requires the isolation of a TLDS, which depends on the symmetry of the system.
For instance, if we choose to start in the $|\pm 1\rangle$ levels fixing the detuning accordingly ($\Delta = \pm \delta$), as shown in Fig.~\ref{Fig1}(b), then we observe the regular Rabi oscillations that basically depend on the pulse area. 
This is because the populated dressed states, $\lambda_1(t)$ and $\lambda_3(t)$,
 are the lowest pair (see Fig.~\ref{Fig1}(d)), and the generalized pulse area, 
${\cal A} = \int_0^T \!dt \left[\lambda_1(t) - \lambda_3(t) \right]$, 
is dominated by the dynamic phase accumulated by the lowest dressed state, 
whose Autler-Townes splitting approximately follows the field envelope. In the adiabatic limit we can obtain analytic results for the asymmetric configuration although the expressions are more cumbersome and we are not presenting them here. In Fig.~\ref{Fig2}(c) we compare the analytical results with the exact numerical solution of the TDSE. The full coincidence of the population shows that the nonadiabatic couplings are totally negligible for the asymmetric configuration. Figure~\ref{Fig2}(d) shows the density plot of the final yield of population transfer as a function of both pulse area, $S_0$,  and splitting,  $\delta$, calculated numerically from the TDSE.

\begin{figure}[htpb!]
\includegraphics[width=7.5cm]{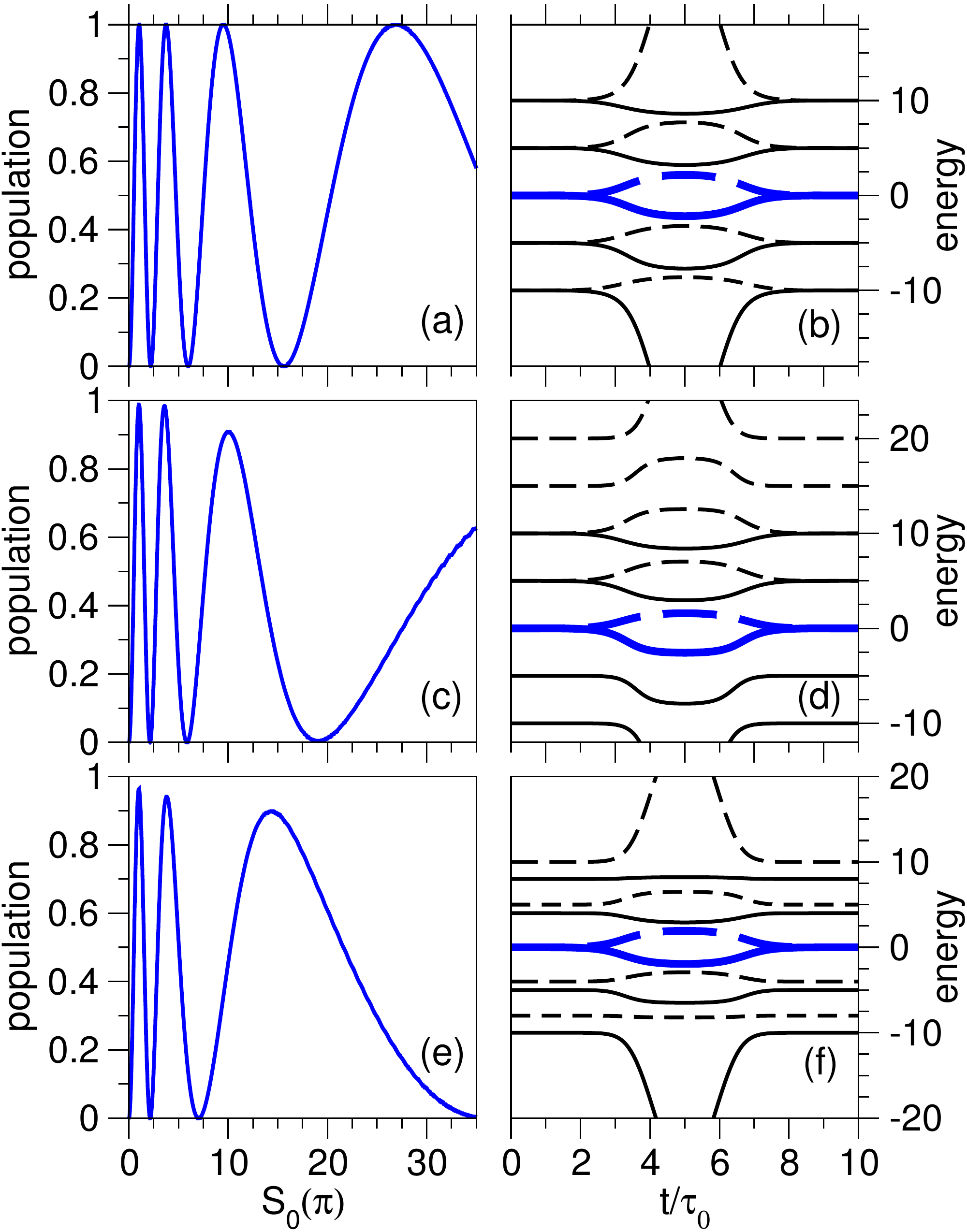}
\caption{Target population at final time as a function of the pulse area (left column) and
the corresponding dressed states for $\Omega_0\tau_0 = 10$ (right column) for a system
of two coupled $5$ level ladders with $\delta\tau_0 = 5$. In (a) and (b) we start and end in the middle of the ladder ($M = M' = 0$). In (c) and (d) we start in the middle and end at the lowest level of the 
ladder ($M'=-2$). In (e) and (f) the target ladder has a different energy splitting ($\delta' = 4$). The populated dressed states are always shown with thicker blue lines.}
\label{Fig3}
\end{figure}

What is essentially required to observe ARO? As mentioned above, we need to isolate the TLDS among the set of all levels. This implies using relatively long pulses, with bandwidth smaller than the energy spacing, such that there is no transient absorption to other states induced by the lack of energy resolution.
The ARO can only be observed at large Rabi frequencies, where the effective Rabi frequency no longer depends on the pulse amplitude, as in the standard Rabi oscillation regime, but on the energy splittings.
A minimum number of $4$ coupled levels are needed, as in the tripod model, but more levels can be involved as long as the initial and the target state are not both the highest or the lowest energy eigenstates of the ground or excited manifolds.

To generalize the ARO demonstrated above for the tripod $(3+1)$-level model we now consider the population transfer in a system of two fully coupled $5$-level ladders, or $5+5$-level system, with Hamiltonian
\begin{equation}
{\sf H} = \sum_M \delta M |M\rangle \langle M| + 
\sum_{M'} (\delta' M' + \Delta) |M'\rangle \langle M'| 
- \Omega(t)/2 \left[ \sum_{M,M'} |M\rangle \langle M'| + \mathrm{h.c.} \right] \, ,
\end{equation} 
where $M, M'=\pm2, \pm1, 0$, and $\delta$, $\delta'$ are the energy splittings of the ground and excited manifolds. The left column of Fig.~\ref{Fig3} shows the population of the target state as a function of the pulse area when we start in the $|0\rangle$ state to reach the target state $|0'\rangle$ using $\Delta = 0$ or the target state $|-2'\rangle$ using $\Delta = -2\delta'$. If the transition is not resonant, an additional detuning modulates the oscillation and affects the maximum population transfer that can be achieved, as in Eq.(1). When the pulse duration is of the order of the period associated with the energy splitting, the population flow is not fully selective but one can still observe ARO between the overall population of the manifolds. For shorter pulses we observe normal Rabi oscillations and lower yields, due to Raman transitions.

The right column of Fig.~\ref{Fig3} shows the corresponding dressed states when $\Omega_0\tau_0 = 10$. The symmetric arrangement minimizes the nonadiabatic couplings isolating the TLDS and a perfect ARO is observed, Fig.~\ref{Fig3}(a),(b). Asymmetries cause larger couplings that lead to some distortions in the ARO and losses in the yield of population inversion, although the anomalous Rabi oscillations can still be observed, Fig.~\ref{Fig3}(c),(d). A similar distortion effect is observed if the Rabi frequencies among the different levels are not equal. Figure~\ref{Fig3}(e),(f) shows the change in the ARO when energy splitting in the manifolds are different. In general, the ARO can be observed when the populated dressed states are constrained by avoided crossing with the near-lying dressed states and this will occur as long as there are no strong selection rules that forbid most couplings, generating block-diagonal Hamiltonians, in which case the TLDS is no longer protected. It is instructive to mention a couple of examples where the ARO cannot be observed. First, the selection rules $\Delta M = \pm 1$ for higher angular momentum states ($M \ge 2$) will forbid the ARO due to breaking necessary coupling arrangement. Second, if the dipole couplings decrease quickly in the manifold as $\Delta M$ increases, then the non-adiabatic couplings cannot be neglected and the pattern of the ARO become more complex. 

In summary, we have shown a new generic effect of coherent excitation of quantum systems. Under strong excitation with long pulses (areas typically larger than $2\pi\delta$) two-level dressed states are isolated by avoided crossing with the remaining dressed states of the system. When the initial or target states are embedded in a manifold of levels, the Rabi oscillation frequency of the populations depend on the characteristic energy splitting and not on the pulse area. This guarantees a much more robust
quantum state preparation that may compete, under certain conditions, with adiabatic passage and could
be useful in quantum information processes. In addition, the sensitivity of the ARO to the energy
splittings could in principle be used to obtain the parameters of the Hamiltonian even when the additional levels in the manifold are never (or very weakly) populated at final time. Thus, coherent population dynamics via the ARO could be used as a spectroscopic method.

This work was supported by the Basic Science Research Program (NRF) funded by Ministry of Science, ICT \& Future of Korea (2017R1A2B1010215) and by the Spanish government through the MICINN project CTQ2015-65033-P.

{}

\end{document}